\documentclass[a4paper,11pt]{article}
\usepackage{bbold}
\usepackage[x11names]{xcolor}
\usepackage{pos}
\usepackage{tcolorbox}

\definecolor{pos}{RGB}{102, 156, 22}

\title{Training School Lecture Notes: WISPs in gamma-ray astrophysics}

\author*[a]{Francesca Calore}
\author[b]{Christopher Eckner}

\affiliation[a]{LAPTh, USMB, CNRS,  F-74940 Annecy, France}

\affiliation[b]{Center for Astrophysics and Cosmology, University of Nova Gorica, \\ Vipavska 11c, 5270 Ajdov\v{s}\v{c}ina, Slovenia}

\emailAdd{calore@lapth.cnrs.fr}
\emailAdd{ceckner@ung.si}

\abstract{
These lecture notes provide an overview of high-energy astrophysical processes involving axions, axion-like particles (ALPs), and other weakly interacting slim particles (WISPs) focusing on their potential observational signatures in astrophysical environments. After introducing key concepts in high-energy astrophysics, we present the fundamental properties of WISPs, emphasizing their phenomenological implications. Particular attention is given to ALP-photon conversion in strong magnetic fields and the possible decay signatures of ALPs in sources such as active galactic nuclei, galaxy clusters, and cosmic-ray accelerators. These effects can lead to distinctive modifications in astrophysical spectra, spatial distributions, and polarization patterns, providing unique probes of physics beyond the Standard Model. We discuss their role in dark matter scenarios and their potential impact on high-energy observations. The lecture series is supplemented by hands-on tutorials, including exercises on axion electrodynamics and an analysis of gamma-ray data from NGC 1275 to search for ALP-photon conversion signatures.}

\FullConference{2nd Training School and General Meeting of the COST Action COSMIC WISPers  (CA21106) (COSMICWISPers2024)\\
 10-14 June 2024 and 3-6 September 2024\\
Ljubljana (Slovenia) and Istanbul (Turkey)\\}

\begin{document}
\maketitle

\newpage
\oldTableofcontents

\section{Introduction}
These lecture notes provide an overview of high-energy astrophysical processes involving axions and axion-like particles (ALPs). The aim is to introduce key concepts in high-energy astrophysics, present the fundamental properties of ALPs, and explore their potential observational signatures in astrophysical sources. Particular attention is given to the processes of ALP-photon conversion in strong magnetic fields and the decay signatures of ALPs and other light particles in various astrophysical environments. These phenomena are explored in the context of different astrophysical sources, such as active galactic nuclei (AGN), galaxy clusters, and cosmic-ray sources.

These notes do not cover several related topics that are beyond the scope of this lecture series. Specifically, the deep theoretical motivations for axions and ALPs, including detailed derivations from quantum field theory, are not addressed. The production mechanisms of weakly interacting slim particles (WISPs) in the early universe and their connection to dark matter (DM) are also outside the scope of these notes. Additionally, we do not discuss constraints from stellar evolution (but see~\cite{Caputo:2024oqc}), nor do we cover the constraints imposed by strongly magnetized objects, including pulsars and magnetars.

Axions and other WISPs -- such as dark photons and sterile neutrinos -- have emerged as promising solutions to several outstanding puzzles in fundamental physics and cosmology. These particles are characterized by sub-GeV masses and extremely weak interactions with Standard Model (SM) particles, often through portal interactions mediated by (pseudo)scalars, vectors, or fermions. 
WISPs, including axions and axion-like particles, provide viable candidates for DM, particularly in the form of cold, non-thermal relics or ultra-light bosons.
Some WISPs, such as sterile neutrinos, are linked to mechanisms like leptogenesis, which can explain why the universe is dominated by matter over antimatter.
Finally, as we will see below, the axion specifically addresses the strong CP problem by dynamically setting the CP-violating term in quantum chromodynamics (QCD) to near zero. The WISPs' weak couplings make them difficult to detect in laboratory experiments, but their effects may manifest in astrophysical and cosmological observations, making high-energy astrophysics a key probe for their discovery.

In these lecture notes, we focus on the \textit{signatures of light, weakly coupled particles} within the realm of \textit{high-energy (HE) astrophysics}. This field covers emissions from astrophysical sources across a wide energy range -- from X rays (keV) to gamma rays (PeV) -- as well as astrophysical diffuse backgrounds observed across multiple wavelengths.
A \textit{signature is a sizeable modification of expected astrophysical signals}, which can manifest as, for e.g.~changes in the spectral distribution, alterations in the spatial distribution, and/or modifications in polarization.  
These deviations from standard astrophysical expectations serve as crucial probes for new physics beyond the SM. In particular, we will consider ALPs as a benchmark model to be scrutinized, and we will comment on other WISPs occasionally.

%
The lecture notes are organized as follows. In Sec.~\ref{sec:introAstro} we provide a brief primer on high-energy astrophysics covering the essential processes that produce non-thermal emission in Galactic and extragalactic environments. The following Sec.~\ref{sec:introPheno} elaborates on the phenomenology of WISPs regarding their interplay with photons while the subsequent Sec.~\ref{sec:astro-HEP-ALP-signals} dives deeper into the astrophysical signatures arising from the coupling of ALPs with photons. The lecture part of this article is concluded in Sec.~\ref{sec:darkmatter} by emphasizing the implications of WISPs as a potential candidate for the DM in the universe. Lastly, we present in Sec.~\ref{sec:tutorials} the series of hands-on exercises supplementing the topics discussed in the lecture. The tutorials cover theoretical problems about axion electrodynamics as well as a data analysis part dissecting gamma-ray data of NGC 1275 for a signature of ALP-photon conversions.

\section{A short introduction to high-energy astrophysics}
\label{sec:introAstro}
The local universe is populated by a large variety of astrophysical objects, mostly hosted by galaxies and clusters of galaxies. 
These objects include, for instance, Active Galactic Nuclei (AGN), Supernovae (SNe), star-forming galaxies as well as the media hosted in galaxy clusters. 
The environments of these objects are rich in magnetic fields and energetic particles, making them ideal candidates for studying axions or ALPs and other exotica. In addition to discrete sources, the universe is permeated by several cosmic photon backgrounds
 that provide further avenues to explore axion-related physics.

The observation of individual, multi-wavelength emitting, astrophysical objects helps us understand the acceleration mechanisms in a variety of extreme environments, as well as in building population models for these emitters.
Here, we review the main classes of high-energy astrophysical sources, highlighting the characteristics that make them suitable for studying axions and the like.

{\bf Active galactic nuclei.} AGN are powered by supermassive black holes that accrete matter from their surroundings, for a review we refer the reader to~\cite{2008NewAR..52..227T}. 
A significant fraction of the total luminosity of an AGN is non-thermal, resulting from two key components:
(i) Emission from the accretion disc surrounding the black hole, and 
(ii) emission from highly collimated relativistic jets extending from sub-parsec to kiloparsec scales.
The magnetic fields within AGN jets have a complex structure, consisting of both poloidal (aligned with the jet axis) and toroidal (perpendicular to the axis) components~\cite{2021Galax...9...58G}. 
The strength of the magnetic field in these jets can reach $B \sim 10^3$ G. 
The coherence length of these fields ranges from sub-parsec near the central black hole to 
several parsecs along the jet's extension. These magnetic fields are essential for synchrotron radiation and could play a role in axion-photon interactions through the (inverse) Primakoff effect (which will be addressed later in Sec.~\ref{sec:introPheno}).

{\bf Supernovae and their remnants.}
SNe are violent explosions marking the end of a star's life, see e.g.~\cite{Orlando:2023alz}. They come in two main types:
(i) Type I SNe result from the accretion of matter onto a white dwarf in a binary system, leading to a thermonuclear runaway, while (ii) Type II SNe occur from the gravitational collapse of the core of a
massive star ($M \gtrsim 8 \, M_{\odot}$).
The explosion expels the outer layers of the star, forming a supernova remnant (SNR) that expands into and shocks the surrounding interstellar medium (ISM). 
The magnetic fields in these remnants reach strengths ranging from $25 \, \mu$G to $1000 \, \mu$G~\cite{2012SSRv..166..231R}. SNe are powerful factories of axions and alike. Besides, shocks in SNRs are ideal sites
 for cosmic-ray acceleration via diffusive shock acceleration, and the magnetic fields also provide a medium for potential axion-photon conversion.

{\bf Star-forming galaxies.}
Star-forming galaxies are characterised by high rates of star formation, particularly in regions known as star-forming or starburst regions. 
These regions are sites of intense stellar activity and can host large-scale magnetic fields ranging from a few $\mu$G to hundreds of $\mu$G, see e.g.~\cite{2006ApJ...645..186T}. The strong 
magnetic fields are thought to be generated by dynamo processes, and they influence the transport of cosmic rays as well as photon-axion mixing in these environments.

{\bf Galaxy clusters.}
Galaxy clusters are the largest gravitationally bound structures in the universe, containing hundreds to thousands of galaxies. 
The intracluster medium (ICM) is composed of hot, X-ray-emitting, highly ionized gas with temperatures reaching $T \sim 10^7$ to $10^8$ K. 
The magnetic fields in galaxy clusters are relatively weak, typically ranging from $0.1 \, \mu$G to a few $\mu$G, for a review see~\cite{Govoni:2004as}.
These magnetic fields are often chaotic and tangled due to the turbulent nature of the ICM, although large-scale ordered fields can also exist. 
The coherence lengths of magnetic fields in galaxy clusters range from 10 to 100 kpc. 

Besides individually resolved objects, the universe is also filled with various cosmic photon backgrounds, originating from different astrophysical processes. 
These backgrounds are mostly of extragalactic origin, formed from a superposition of faint photon emitters or truly diffuse processes -- mostly from cosmic-ray interactions. The cosmic photon background extends from radio wavelengths up to $\mathcal{O}(100)$ GeV. The main components at low energies are the Cosmic Microwave Background (CMB), a relic of the Big Bang, with a temperature of 2.7255 K after the subtraction of Galactic foregrounds, the 
Cosmic Infrared Background (CIB), which represents the emission from dust heated by stars in unresolved galaxies,
the Cosmic Optical Background (COB), mainly due to starlight from galaxies and difficult to clean from Galactic contamination, and the 
Cosmic Ultraviolet Background (CUB), which originates from ionizing sources such as star-forming galaxies and quasars.
At the highest energies, the Cosmic X-ray Background (CXB) is believed to be dominated by bremsstrahlung radiation from hot accretion discs around AGN, and the Cosmic Gamma-ray Background (CGB) originates from the superposition of large populations of AGNs, star-forming galaxies, and galactic sources. For a dedicated review of these cosmic photon backgrounds, we refer the reader to~\cite{Hill:2018trh}.
Each of these backgrounds offers a unique probe of the large-scale structure of the universe and the potential interactions between photons and axions in extragalactic environments.

At higher energies, diffuse gamma-ray emission has been detected by ground-based telescopes such as the High Energy Spectroscopic System (H.E.S.S.)~\cite{HESS:2014ree, HESS:2017tce}, the High-Altitude Water Cherenkov Gamma-Ray Observatory (HAWC)~\cite{HAWC:2021bvb, HAWC:2023wdq}, the Tibet AS$\gamma$ experiment~\cite{2021PhRvL.126n1101A}, and the Large High Altitude Air Shower Observatory (LHAASO)~\cite{LHAASO:2023gne, LHAASO:2024lnz}.
This emission is primarily of Galactic origin, constrained by the horizon of very high-energy (VHE) photons, which are absorbed through pair production interactions with the extragalactic background light (EBL), made up
by the CIB, COB, and CUB~\cite{Franceschini:2021wkr}.

\section{Basics of WISP phenomenology}
\label{sec:introPheno}
This section covers the key concepts of WISP phenomenology, with a primary focus on their coupling with photons. We will present the two-photon decay process and, being it the central process for ALP phenomenology, the ALP-photon mixing. A similar process applies to dark photons, where mixing with photons also governs their interactions and effects in various astrophysical and cosmological contexts~(see, e.g., \cite{Fabbrichesi:2020wbt}).

\subsection{Spontaneous particle decay}
\label{sec:introPheno-decay}

Spontaneous decay processes are particularly relevant for low-mass particles, which are often suppressed through loop-level interactions. In general, BSM  particles can decay spontaneously into two lighter particles, a process that can be described by a specific decay rate, $\Gamma$, and lifetime, $\tau$. The decay products, such as photons, are emitted back-to-back in the rest frame of the decaying particle, with each photon carrying half of the total energy of the decaying particle.
For a particle at rest, such as particles forming the DM halo of the Milky Way, the decay results in a sharp photon line with energy strongly correlated with the ALP mass. However, when the ALP is boosted, or when it has a specific spectral energy distribution, the resulting signal broadens, and the precise spectral shape depends on the ALP production process.

For heavier ALPs, particularly those with masses greater than a few keV, spontaneous decay into photons becomes an increasingly significant production channel. While photon-ALP conversion is typically suppressed in this mass range, the decay rate remains non-negligible. The decay rate for ALPs can be written as~\cite{Higaki:2014zua}:
\begin{equation}
     \Gamma(a \rightarrow \gamma \gamma)=\frac{g_{a\gamma}^2m_a^3}{64\pi} =0.755\times 10^{-30}\left(\frac{g_{a\gamma}}{10^{-20} \, \rm{GeV}^{-1}}\right)^{2}\left(\frac{m_a}{100 \, \rm{keV}} \right)^3\, {\rm s}^{-1}\,,
\end{equation}
where $g_{a\gamma}$ is the ALP-photon coupling and $m_a$ the ALPs mass.

In addition to spontaneous decay, stimulated axion decay has been studied in the literature \cite{Caputo:2018vmy, Caputo:2018ljp}, providing a deeper understanding of ALP decay processes under specific conditions.

Another example of decaying WISPs is the sterile neutrino~\citep{Drewes:2016upu}, which can decay into one photon 
and one active neutrino via a loop-mediated radiative process. 
The width for this channel is given by (e.g.~\cite{Bezrukov:2009th})
\begin{equation}
    \Gamma(N \rightarrow \nu\gamma) \simeq \frac{9\alpha_{\mathrm{em}} G_F^2 m_N^5\sin^2(2\theta)}{1024\pi^4} \simeq 1.36 \times 10^{-29} ~ {\rm s}^{-1} \left[ \frac{ \sin^2(2\theta)}{10^{-7}} \right]\left(\frac{m_{N}}{1~ {\rm keV}} \right)^5,\label{gammanuwidth}
\end{equation}
where $\alpha_{\mathrm{em}}$ denotes the fine-structure constant, $G_F$ is the Fermi constant, $\theta$ refers to the mixing angle between the sterile and active neutrino, and $m_N$ is the sterile neutrino mass. 

If these particles constitute a portion of DM, their decay could lead to detectable photon emissions from distant galaxies or galaxy clusters, or any regions characterized by high DM densities, cf.~Sec.~\ref{sec:darkmatter}.

ALPs produced in stellar events, such as SNe, can also decay into photons, leading to observable signals. For core-collapse SNe (cc-SNe), one can look for time-dependent gamma-ray bursts or the cumulative emission from these decays over time and redshift. One of the strongest constraints on MeV-mass ALPs comes from the non-observation of gamma-ray decay signals from SN1987A~\cite{Muller:2023pip}, along with limits from low-energy SNe \cite{Caputo:2022mah} and the contribution of ALPs from past cc-SNe to the extragalactic diffuse gamma-ray emission~\cite{Calore:2020tjw, Caputo:2021rux, Caputo:2022mah}.

Beyond ALP-photon coupling, other couplings leading to decay have been studied, in particular, couplings with electrons for generic feebly interacting particles produced in SNe and decaying into electron-positron pairs. Examples include~\cite{Calore:2021lih, DelaTorreLuque:2024zsr, Carenza:2023old}, which explore such processes in stellar environments. 

\subsection{Fundamentals of ALP-photon mixing}

We will explore the ALP-photon mixing in more detail throughout this section. The photon-ALP mixing, driven by the Primakoff process, can be derived from the corresponding Lagrangian~\cite{Raffelt:1987im}
\begin{equation}
\label{eq:ALP_lagrangian}
\mathcal{L}_{a\gamma}=-\frac{1}{4} g_{a\gamma} F_{\mu\nu}\tilde{F}^{\mu\nu}a=g_{a\gamma} {\bf E}\cdot{\bf B}\,a \, ,
\end{equation}
which results in the well-known conversion of ALPs into photons when an external magnetic field ${\bf B}$ is present.
Here, we summarise the key equations and components required to calculate the probability of photon-ALP conversions. For a more detailed discussion of the theoretical and mathematical aspects, we refer to~\cite{Raffelt:1987im, DeAngelis:2007dqd, Mirizzi:2009aj, DeAngelis:2011id, Meyer:2014epa, Kartavtsev:2016doq}.

Photon-ALP conversion requires the presence of an external magnetic field \(\mathbf{B}\), with a non-zero transverse component \(\mathbf{B}_{\perp}\) relative to the propagation direction. If the initial state propagates along \(\hat{\mathbf{z}}\), and \(\mathbf{B}_{\perp}\) is given by \(B(\cos\theta, \sin\theta, 0)^T\), where \(\theta\) represents the angle between \(\mathbf{B}_{\perp}\) and the \(\hat{\mathbf{x}}\)-axis, the photon polarisation states \(A_x\) and \(A_y\) combine to form two components: \(A_{\perp}\), perpendicular to \(\mathbf{B}_{\perp}\), and \(A_{\parallel}\), parallel to \(\mathbf{B}_{\perp}\). Only the \(A_{\parallel}\) component is capable of converting into ALP states.

The evolution of a photon-ALP state propagating along \(\hat{\mathbf{z}}\) is governed by~\cite{Raffelt:1987im, Kartavtsev:2016doq}:
\begin{equation}
\label{eq:ALP-photon-prop-pure}
i\frac{\mathrm{d}\mathcal{A}}{\mathrm{d} z} = \left(\mathcal{H}_{\mathrm{dis}} - \frac{i}{2} \mathcal{H}_{\mathrm{abs}}\right) \mathcal{A},
\end{equation}
where $\mathcal{A} = \left(A_{\perp}, A_{\parallel}, a\right)^T$ represents the photon polarisation states and the ALP state $a$. 
The dispersion Hamiltonian $\mathcal{H}_{\mathrm{dis}}$ expresses photon-ALP mixing and writes as~\cite{Raffelt:1987im, Mirizzi:2009aj}:
\begin{equation}
\label{eq:mixing_matrix}
\mathcal{H}_{\mathrm{dis}} = \begin{pmatrix}
\Delta_{\perp} & 0 & 0 \\
0 & \Delta_{\parallel} & \Delta_{a\gamma} \\
0 & \Delta_{a\gamma} & \Delta_{a}
\end{pmatrix},
\end{equation}
with components defined as~\cite{Mirizzi:2009aj, Kartavtsev:2016doq}:
\begin{align}
\Delta_{\perp} &= \Delta_{\mathrm{pl}} + 2\Delta_{\mathrm{B}} + \Delta_{\gamma\gamma}, \\
\Delta_{\parallel} &= \Delta_{\mathrm{pl}} + \frac{7}{2}\Delta_{\mathrm{B}} + \Delta_{\gamma\gamma}, \\
\Delta_{a} &= -\frac{m_a^2}{2\omega}, \\
\Delta_{a\gamma} &= \frac{g_{a\gamma}}{2}B.
\end{align}
Here, 
the dispersion due to plasma density, magnetic field, and photon-photon interactions, are described by $\Delta_{\mathrm{pl}}$, $\Delta_{\mathrm{B}}$, and $\Delta_{\gamma\gamma}$, respectively~\cite{Dobrynina:2014qba}.
The absorption Hamiltonian $\mathcal{H}_{\mathrm{abs}}$ accounts for photon absorption effects and is expressed as~\cite{Kartavtsev:2016doq}:
\begin{equation}
\label{eq:absorption_matrix}
\mathcal{H}_{\mathrm{abs}} = \begin{pmatrix}
\Gamma & 0 & 0 \\
0 & \Gamma & 0 \\
0 & 0 & 0
\end{pmatrix},
\end{equation}
where $\Gamma$ represents photon-photon absorption~\cite{Mirizzi:2009aj}. The
particular environment's conditions set the 
importance 
of the absorption term, which can describe e.g.~Galactic absorption and/or absorption on the CMB and EBL.

In a homogeneous or slowly varying $B$-field, a photon beam
develops a coherent axion component.
In the limit of a purely polarised photon beam in a single magnetic domain $L$ with a coherent $B$-field, the propagation equations reduce to a 2-dimensional problem, where one can define~\cite{Mirizzi:2009aj}:
\begin{equation}
    \frac{1}{2} \tan( 2\theta )= \frac{\Delta_{a \gamma}}{\Delta_{\parallel} - \Delta_a }  \, ,
\end{equation}
as the angle that diagonalises the mixing matrix. 
Similarly to neutrino oscillations then, the probability for a purely polarised photon beam to oscillate into an ALP after distance $L$ 
writes as:
\begin{equation}
    P(a \rightarrow \gamma) =\sin^2(2 \theta) \sin^2 (\Delta_{\rm osc} L /2 )    \, , 
\end{equation}
where $\Delta_{\rm osc} \equiv [(\Delta_\parallel - \Delta_a)^2 + 4 \Delta_{a\gamma}^2)]^{1/2}$ (see also Sec.~\ref{sec:tutorial_ex2} for a pedagogic derivation in a slightly simplified setting).
Therefore, the mixing is maximum when $\Delta_{a\gamma} \gg \left|\Delta_\parallel - \Delta_a\right|$ (strong mixing regime). On the other hand, the probability is suppressed by plasma or CMB effects when $\Delta_{a\gamma} \ll \left|\Delta_\parallel - \Delta_a\right|$.
In the limit of massless ALPs, sufficiently low couplings and low energies, the probability simplifies to: 
\begin{equation}
    P(a \rightarrow \gamma) = \Delta_{a \gamma}^2 \frac{\sin^2 (\Delta_{\rm osc} L /2 )}{(\Delta_{\rm osc} L /2 )^2}    \, .
\end{equation}

\begin{figure}
    \centering
    \includegraphics[width=0.5\linewidth]{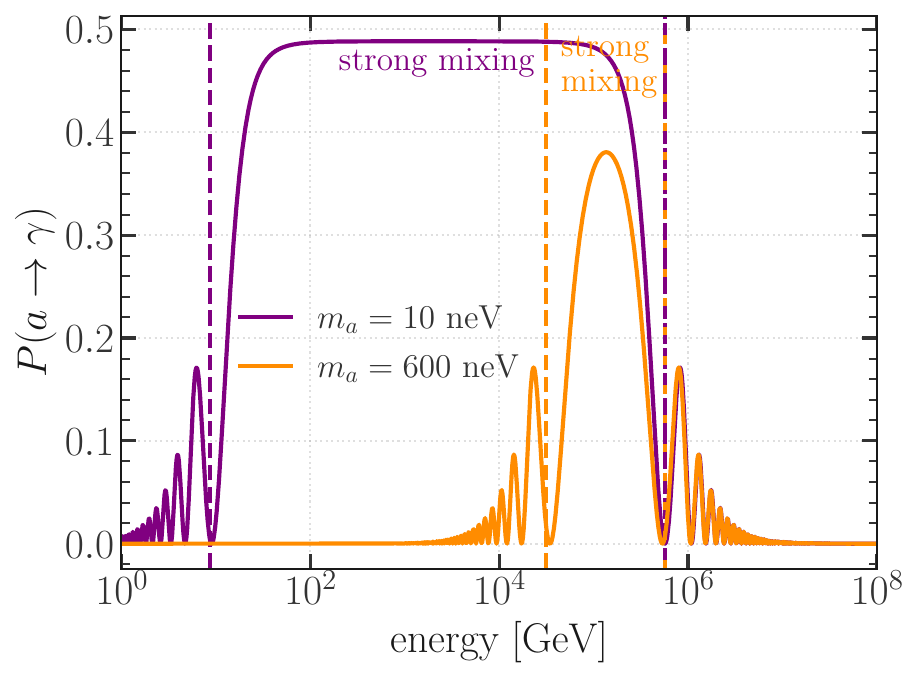}
    \caption{Dependence of the ALP-photon conversion probability $P(a \rightarrow \gamma)$ on the ALP mass. We display the cases $m_a = 10$ neV (purple) and $m_a = 600$ neV (orange) for a coupling strength of $g_{a\gamma} = 3\times10^{-11}$ GeV$^{-1}$. The dashed vertical lines mark the onset and end of the strong mixing regime with energy. Note that the upper boundary of this regime is the same for both ALP masses. The conversion probability was calculated for a single domain featuring a coherent magnetic field with parameters as defined in the tutorial of the ALP propagation software \texttt{gammaALPs} \cite{Meyer:2021pbp}.}
    \label{fig:Pag}
\end{figure}

The oscillation length is $L_{\rm osc} = 2\pi/\Delta_{\rm osc}$, with coherent oscillations in the case $L \ll L_{\rm osc}$.
In this case, the dominating plasma term can be expressed as: 
\begin{equation}
    \Delta_{\rm osc} = \left[\frac{(m_a^2 - \omega_{\rm pl}^2)^2}{4 \omega^2} + (g_{a\gamma} B)^2 \right]^{1/2} \, ,
\end{equation}
with $\omega_{\rm pl}^2 \equiv (4 \pi \alpha_{\rm em} n_e)/m_e$ determined by the environmental, 
free electron density $n_e$.
We can then write:
\begin{equation}
    \Delta_{\rm osc} = 2 \Delta_{a\gamma} \sqrt{1 + \frac{\omega_c^2}{\omega^2}} \, ,
\end{equation}
where the critical energy, $\omega_c$, is defined as:
\begin{equation}
    \omega_c = 2.5\,   \mathrm{GeV} \Bigl( \frac{m_{\it a}^2 - \omega_{\mathrm{pl}}^2}{1 \, \rm neV^2} \Bigr) \Bigl(\frac{1 \, \rm \mu G}{B}\Bigr) \Bigl(\frac{10^{-11} \, \rm GeV^{-1}}{g_{{\it a}\gamma}}\Bigr) \, .
\end{equation}
The critical energy varies depending on the ALP mass and free electron density, and it takes values of $\sim$ GeV for neV masses and $\sim$ keV for $10^{-12}$ eV masses.
This critical energy differentiates two regimes: (a) the strong mixing regime ($\omega \gg \omega_c$), where the conversion probability is approximated as $P(a \rightarrow \gamma)\propto g_{a\gamma}^2 B^2 L^2$, and it is therefore energy independent; (b) the oscillation regime ($\omega \simeq \omega_c$) where the conversion probability possesses an oscillatory, energy-dependent, behaviour.  
In Fig.~\ref{fig:Pag}, we show the conversion probability over a broad range in energy to exemplify the dependence of the oscillatory pattern on the ALP mass (here: $m_a = 10$ neV and 600 neV) at a fixed value of $g_{a\gamma} = 3\times10^{-11}$ GeV$^{-1}$.
For a study of the conversion probability as a function of ALP mass, we refer the reader to~\cite{Calore:2023srn}.

For more complex scenarios, one can use the density matrix formalism. The density matrix $\bf{\rho}$, defined as $\bf{\rho} = \mathcal{A}\otimes\mathcal{A}^{\dagger}$, evolution is described as in~\cite{Mirizzi:2009aj, Kartavtsev:2016doq}:
\begin{equation}
\label{eq:ALP-photon-prop-density}
i\frac{\mathrm{d}\bf{\rho}}{\mathrm{d} z} = \left[ \mathcal{H}_{\mathrm{dis}}, \bf{\rho} \right] - \frac{i}{2} \{\mathcal{H}_{\mathrm{abs}}, \bf{\rho}\}.
\end{equation}

The next section is dedicated to the 
the signatures of ALP-photon mixing in high-energy astrophysics.

\begin{tcolorbox}[colback=pos!5!white, colframe=pos!75!black, title=Exercise 1]
Derivation of Maxwell’s equations for axion electrodynamics, cf.~\ref{sec:tutorial_ex1} for full details.
\end{tcolorbox}

\begin{tcolorbox}[colback=pos!5!white, colframe=pos!75!black, title=Exercise 2]
 Derivation of the conversion probability in a constant magnetic field, cf.~\ref{sec:tutorial_ex2} for full details.
\end{tcolorbox}

\section{High-energy astrophysical signatures of ALP-photon mixing}
\label{sec:astro-HEP-ALP-signals}

\subsection{Main astrophysical observables}
The search for WISPs, particularly axions and ALPs, in astrophysical environments relies on key observables that provide indirect evidence of their existence. These observables include the spectral energy distribution (SED) of individual sources, the time-dependence of the photon flux, and the diffuse large-scale flux.

The SED represents the energy output of a source across different wavelengths, measured in $\rm erg/cm^2/s$, and provides detailed information about physical processes such as synchrotron radiation and inverse Compton scattering. Deviations from the expected SED can indicate axion-photon conversion. In addition to the SED, the time variability of the photon flux is important, as changes in photon output over time can indicate potential axion-photon conversions. These time-dependent observations are vital for detecting transient or variable events that could be associated with axions.
The diffuse large-scale flux offers spatial information about photon emission across the sky, averaged over a region of interest and measured in $1/\rm sr$. This flux arises from unresolved sources or truly diffuse processes, forming the cosmic radiation background, as introduced above. Studying this emission helps us understand the distribution of photon sources and their interaction with cosmic magnetic fields.

The cumulative emission of ALPs from astrophysical objects can leave imprints in the diffuse large-scale gamma-ray emission, detectable as brightening or specific spatial patterns. Instruments with a large instantaneous field of view (FoV) and good angular resolution, like the Large Area Telescope (LAT) aboard the \textit{Fermi} satellite~\cite{Fermi-LAT:2009ihh} or ground-based water-Cherenkov telescopes, like HAWC \cite{Abeysekara:2017mjj} and LHAASO  \cite{Aharonian:2020iou, LHAASO:2021zta}, are ideal for detecting these emissions.
On the other hand, individual high-energy sources, such as AGNs, SNe, and galaxy clusters, also provide avenues for ALP searches. The environmental conditions in these sources can lead to ALP production, detectable through the time-integrated SED. Prompt gamma-ray emission from cc-SNe, spectral distortions, and spectral hardening are key signatures. Instruments with a large FoV, good energy resolution, and sensitivities beyond 10 TeV, such as \textit{Fermi}-LAT, imaging atmospheric Cherenkov telescopes, and water-Cherenkov facilities, are suitable for these searches.
Time-dependent SED observations, particularly from violent gamma-ray bursts triggered by SNe, offer more information about ALP production processes and host conditions. Good temporal resolution and short dead time are essential for these searches.

\subsection{Production of ALPs in astrophysical environments}
Axions and similar particles can be produced efficiently in various astrophysical systems through direct mechanisms in stellar cores or photon-ALP conversion in magnetic fields. These processes are significant in extragalactic environments with strong magnetic fields and energetic emissions.

\textbf{Direct production in stars.}
ALPs can be produced in stellar cores. 
In particular, cc-SNe are significant astrophysical sites for the production of ALPs, with the Primakoff effect being the dominant production mechanism for ALPs with masses below the stellar temperature.~\footnote{ALPs with masses above the stellar temperature may also be produced in such environments. In that case, the process of photon coalescence becomes the dominant production channel (see, e.g., \cite{Candon:2024eah} for the production in M82, \cite{Nguyen:2023czp} for main sequence stars, \cite{Carenza:2020zil} for globular clusters, \cite{Lucente:2020whw} and \cite{Caputo:2022mah} for SNe).} In this process, thermal photons from the hot, dense stellar plasma are converted into ALPs through interactions with the electrostatic fields of ions, electrons, and protons within the star.
ALP production primarily occurs during the explosion phase, which lasts around 1000 milliseconds after core collapse. During the early stages -- about 10 seconds after the explosion onset -- a proto-neutron star forms. At this stage, the star's core has extreme densities and temperatures, making it a prime environment for efficient ALP production~\cite{Raffelt:1996wa}.
The Primakoff process is most relevant in environments where strong electrostatic fields are present, such as in main-sequence stars, horizontal branch stars, and white dwarfs, in addition to cc-SNe. In cases where nucleon-nucleon bremsstrahlung is active, it can become the dominant production channel for ALPs in these environments~\cite{Hanhart:2000ae,Carenza:2019pxu,Carenza:2020cis}.
The ALP spectrum produced in a cc-SN typically peaks between 60 and 80 MeV.
This process also contributes to significant ALP rates in external galaxies like M82 and M87~\cite{Nguyen:2023czp, Ning:2024eky, Candon:2024eah}. For a review on ALP production in stellar systems, including electron and nucleon couplings, see~\cite{Carenza:2024ehj}.

\textbf{ALP production in neutron star mergers.}
Neutron star mergers are key extragalactic events for ALP production due to their strong magnetic fields and extreme conditions. These mergers produce electromagnetic radiation bursts and intense magnetic fields, facilitating photon-ALP conversion~\cite{Diamond:2021ekg, Fiorillo:2022piv}. Observational signatures may include anomalies in the electromagnetic spectrum or gravitational wave signals correlated with gamma-ray bursts~\cite{Dev:2023hax, Diamond:2023cto}.

\textbf{In-situ photon-ALP conversion.}
Extragalactic gamma-ray emitters like AGN, SN remnants, and star-forming galaxies are ideal for photon-ALP conversion due to their strong magnetic fields and high-energy photons. These photons convert into ALPs via the Primakoff effect, impacting the source's SED and diffuse backgrounds~\cite{Vogel:2017fmc}. The conversion efficiency is uncertain due to variations in the intergalactic stellar medium, radiation fields, and magnetic field properties.

In galaxy clusters, the conversion of photons into ALPs occurs in turbulent magnetic fields with random configurations. These fields remain constant within distinct magnetic domains, but their orientation varies from one domain to another. To model the conversion probability accurately, it is necessary to perform an ensemble average over all possible field realizations across these domains.  
In the absence of significant photon absorption, the conversion probability can reach a maximum value of 2/3. However, due to the stochastic nature of turbulent fields, the transfer function shows a large variance across different realizations. This introduces considerable uncertainty in the predicted transparency of very HE photons~\cite{Govoni:2004as, Mirizzi:2009aj}.  

For ALP searches, this stochasticity poses a challenge when interpreting the anomalous transparency of TeV photons. While some models predict significant photon survival at TeV energies due to ALP-photon conversion, the variance in the transfer function complicates efforts to distinguish ALP signatures from other astrophysical effects, such as uncertainties in EBL models.
Despite these challenges, a key potential signature of ALPs in clusters is spatial-dependent anomalous transparency, where regions of the sky display enhanced photon transparency due to the specific configuration of turbulent magnetic fields in the cluster. This distinctive feature provides a promising avenue for detecting ALP effects in high-energy astrophysics~\cite{Mirizzi:2009aj}.

\subsection{Production of photons from ALPs and signatures}
ALPs interact with the cosmic environment during their propagation through the Galactic and extragalactic space, leading to photon production across various wavelengths. 

\begin{figure}
    \centering
    \includegraphics[width=0.8\linewidth]{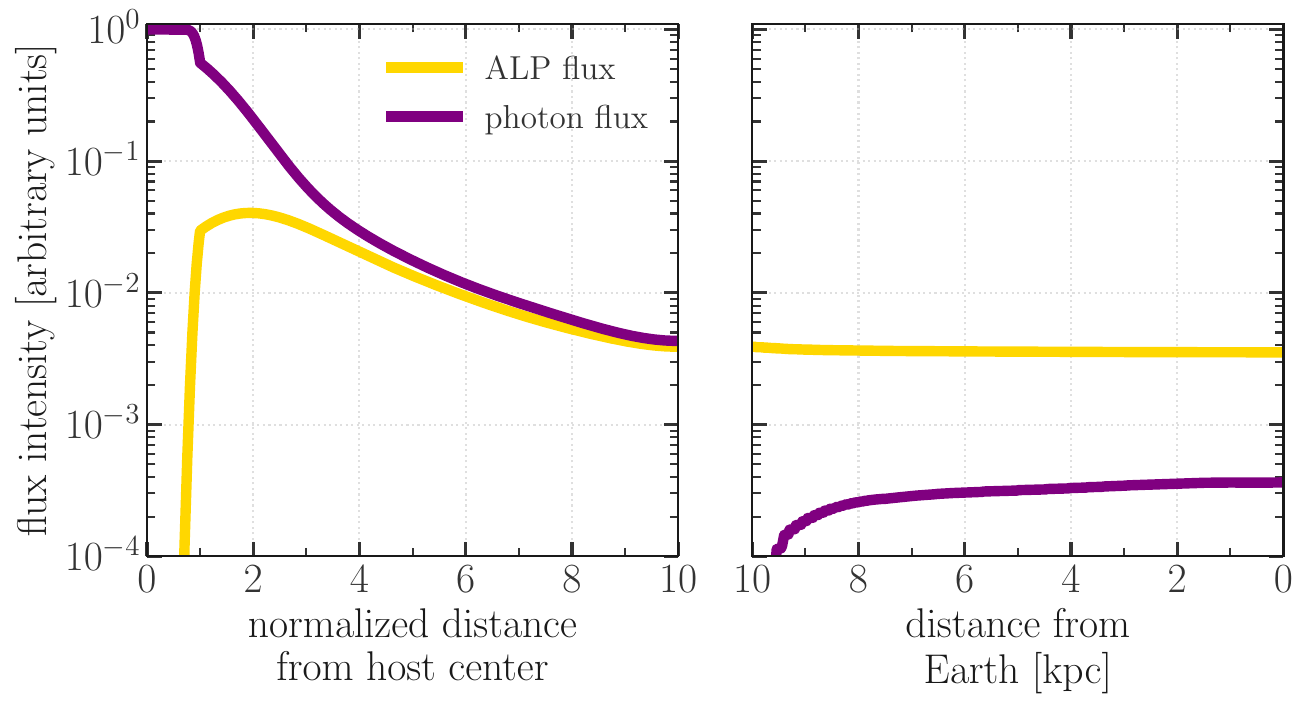}
    \caption{Schematic representation of the evolution of photon (purple) and ALP (yellow) fluxes as a function of distance travelled within a distant extragalactic host source (left panel) or the Milky Way (right panel). The photon flux is normalized to one, with the ALP flux scaled accordingly. The distance in the left panel is given in units of the host's magnetic field coherence length. This panel also illustrates the effect of photon attenuation on the EBL and CMB, which reduces the overall gamma-ray flux beyond the portion converting into ALPs. Both photon and ALP fluxes were calculated using the ALP propagation software \texttt{gammaALPs}. }
    \label{fig:conversion_prob}
\end{figure}

ALPs can convert into photons (and vice versa) in magnetic fields, with the conversion probability depending on ALP properties and the strength and coherence length of the magnetic field. This process occurs in both extragalactic and Galactic magnetic fields, influencing observable photon spectra from distant astrophysical sources.
Magnetic fields in the universe are far from homogeneous, and their impact on ALPs propagation is significantly more complex than idealized scenarios suggest. To properly account for these complexities, several factors must be considered. Fluctuations in the free electron density affect the strength and distribution of magnetic fields along the line of sight, while the stochastic nature of turbulent magnetic components introduces randomness into photon propagation. Additionally, simplified approaches, such as the domain-like approximation, often fail to capture the intricate structure of real magnetic fields. Furthermore, photon absorption at the highest energies must be taken into account to accurately predict observable effects. Overall, realistic models of magnetic field configurations are essential for reliable predictions in the study of high-energy signatures of ALPs.
The intergalactic magnetic field, linked to primordial magnetism, is weak (upper limits of a few $nG$ or $pG$), making the conversion rate negligible~\cite{AlvesBatista:2021sln}. However, conversion in the Galactic magnetic field is significant and depends on the incident line of sight. The right panel of Fig.~\ref{fig:conversion_prob} illustrates how an extragalactic ALP flux partially converts into photons within the Galactic magnetic field, with the yield depending on the direction of the ALP flux due to the Milky Way's magnetic field structure. \footnote{The Milky Way's magnetic field can be studied through several observational techniques. Polarized synchrotron emission from radio to microwave frequencies reveals large-scale field structures in regions where relativistic electrons interact with magnetic fields. Polarized dust emission at sub-millimetre wavelengths traces field lines in dense regions through grain alignment, while starlight polarization shows the field's influence in less dense areas. Faraday rotation, which measures the change in the polarization angle of radio waves through magnetized plasma, provides insights into the field's strength and orientation. Together, these methods offer a multi-scale view of the Galaxy’s magnetic field.
Reconstructing the Milky Way’s magnetic field also depends on the spatial and spectral distributions of cosmic rays. The spatial distribution, or density of cosmic rays across the Galaxy, influences the observed synchrotron intensity, while the spectral distribution, describing their energy spread, affects the frequency dependence of the emission. Both factors are crucial for reliable reconstructions, as uncertainties in these distributions can alter the interpretation of observational data. 
For a comprehensive review, we refer the reader to~\cite{Jaffe:2019iuk}.} 

The Milky Way’s magnetic field consists of both regular and turbulent components, each playing a distinct role in shaping the overall field structure.  
The regular, ordered field includes contributions from the Galactic disc and spiral arms, where field lines generally follow the large-scale structure of the Galaxy. In addition to this disc component, the Galactic halo hosts toroidal fields that wrap around the plane of the disc and poloidal fields that extend vertically above and below it~\cite{Unger:2023lob}.
In contrast, the turbulent, random field has a comparable strength to the regular field but features a much shorter correlation length, typically ranging from 20 to 200 parsecs. This random component arises from turbulent processes in the interstellar medium, significantly affecting local magnetic field variations and contributing to depolarization effects in observational data.
Both regular and turbulent components of the Galactic magnetic field affect the conversion probability, even when the correlation length is smaller than the ALP oscillation length~\cite{Carenza:2021alz}.

\textbf{Signatures.} Photon-ALP conversion leads to various observable signatures from extragalactic sources, distinguishable as signals from individual astrophysical emitters or contributions to cosmic photon backgrounds.

\begin{itemize}
\item \textit{Point-source prompt gamma-ray emission.} In cc-SNe, ALPs can alter the typical gamma-ray emission expectations. ALPs produced in the proto-neutron star matter can escape and convert into photons within the Milky Way or the progenitor's magnetic field.
In particular, if a SN occurs in the Milky Way, the resulting ALP flux would be suppressed only by a factor of $1/d^2$, where $d$ is the distance to the SN. This suppression affects both the ALP flux and the resulting gamma-ray flux from ALP-photon conversion equally.
Moreover, for massless ALPs and feeble couplings, there is a one-to-one correspondence between the ALP energy and the photon energy after conversion, since the conversion probability is energy-independent. 
This results in a prompt gamma-ray burst coincident with the neutrino burst. 
The temporal profile of the gamma-ray signal would mirror the ALP burst duration, providing a distinct transient signature, while the spectral shape could reveal features linked to the underlying ALP production mechanisms and mass~\cite{Payez:2014xsa, Lella:2024hfk, Manzari:2024jns, Raffelt:1996wa}. The photon bursts typically last around 10 seconds with energies peaking at 60 -- 80 MeV~\cite{Calore:2021hhn, Caputo:2022mah}.

\item \textit{Point-source spectral distortions.} 
If ALPs have a non-zero mass,  the conversion probability of ALPs into photons transitions from an energy-independent to an energy-dependent regime, characterized by rapid oscillations that imprint on the SED of gamma-ray sources. The spectrum of the converted photons can therefore exhibit spectral irregularities, introducing features that may vary depending on the ALP mass.
This differs from the smooth spectra of conventional astrophysical sources. 
To effectively search for spectral irregularities, several basic requirements must be met. Firstly, very bright gamma-ray sources are essential as they provide high statistics, which are crucial for accurate spectral determination. Additionally, the sources should be far enough and positioned in the direction of strong transversal B-fields, such as behind or within a galaxy cluster. Good knowledge of the B-field is imperative because ALP searches are sensitive to the product \( g_{a\gamma} B_T \), meaning the constraint on \( g_{a\gamma} \) is only as reliable as the knowledge of \( B_T \).
For Galactic targets, the requirement is simpler, involving only the modelling of the Galactic B-field. The strength of the conversion depends on the position within the Galaxy, for example, beyond the spiral arms. However, there are larger systematics on spectral determination due to the gamma-ray diffuse emission foreground.
Extragalactic targets, on the other hand, require more complex modelling of several B-fields, including intra-cluster, intergalactic, and Galactic fields. The dependence is solely on the latitude and longitude of the sources, and the spectral determination is very accurate. Additionally, modelling of EBL absorption is necessary, which adds further uncertainty.
For a general formalism to look for spectral irregularities, we refer to~\cite{Kachelriess:2023fta, Kachelriess:2022you}.
Spectral distortions due to ALP-photon couplings were looked for at X-ray and gamma-ray energies. 
In gamma rays, an analysis of six bright pulsars using \textit{Fermi}-LAT data revealed a 4.6$\sigma$ significance for common ALP-photon mixing, with spectral variations ranging from 20\% to 40\%, compared to an experimental systematic uncertainty of approximately 3\%~\cite{Majumdar:2018sbv}. Systematic theoretical uncertainties on transverse B field and pulsar distances
were later fully accounted for in~\cite{Pallathadka:2020vwu}.
Studies of three bright SNRs with \textit{Fermi}-LAT and HESS/MAGIC/VERITAS showed a 3$\sigma$ significance for only one source, IC443~\cite{Xia:2019yud}. There are large systematics due to the calibration of GeV-TeV data. An analysis of ten SNRs and pulsars found no evidence for common ALP-photon mixing~\cite{Liang:2018mqm}.
Gamma-ray searches from extragalactic targets showed no evidence for ALP-photon mixing, with strong but not very robust upper limits.
An analysis of the radio galaxy NGC 1275 in the Perseus cluster using \textit{Fermi}-LAT and MAGIC data found that limits are very sensitive to the modelling of the intra-cluster B-field~\cite{Fermi-LAT:2016nkz}. Typically, only the turbulent component is modelled, but there is evidence for a large-scale ordered component, which better matches Faraday rotation measures and other observations. With a purely ordered B-field, the limits almost vanish~\cite{Libanov:2019fzq}.
An analysis of the nearby blazar PKS 2155-304 with \textit{Fermi}-LAT and HESS considered only the turbulent component of the intra-cluster B-field~\cite{HESS:2013udx, Zhang:2018wpc, Guo:2020kiq}. The intergalactic B-field root-mean-square is usually overestimated, which can significantly weaken the limits~\cite{Jedamzik:2018itu}.

Searches for X-ray irregularities in various sources, including the Hydra A galaxy cluster (redshift $z = 0.052$, 240ks \textit{Chandra} observation, \cite{Wouters:2013hua}), NGC 1275 in the Perseus cluster (\textit{Chandra} and \textit{XMM-Newton} observations, \cite{Berg:2016ese, Reynolds:2019uqt}), M87 AGN in the Virgo cluster~\cite{Marsh:2017yvc}, and seven quasars/AGN behind or within nearby clusters (\textit{Chandra} archival data, \cite{Conlon:2017qcw}), found that all spectra are consistent with absorbed power laws. X-rays show strong emission in the low-mass regime, around \( m_a \approx 10^{-12} \) eV, where the ALP mass is below the plasma frequency of galaxy clusters.

\item \textit{Point-source spectral hardening.} ALP-induced transparency at very HE can harden the spectra of distant gamma-ray sources, as HE photons converting into ALPs travel further before reconverting into detectable photons. This enhances the universe's transparency to very HE photons, providing an indirect ALP signature~\cite{Biteau:2015xpa, Dominguez:2011xy, Horns:2012kw, Montanino:2017ara}. The effect depends on magnetic field configurations and ALP properties.
Current searches focus on detecting any anomalous EBL absorption that could be induced by ALPs.
In~\cite{Buehler:2020qsn}, the potential imprint of ALPs in the highest-energy photons emitted by hard gamma-ray blazars is investigated. Despite these efforts, there is currently no evidence suggesting an increased gamma-ray transparency due to ALPs. The distribution of observed highest-energy photons from sources at redshift \( z \geq 0.1 \) aligns with theoretical predictions that do not include the presence of ALPs. This alignment indicates that the current data do not support the hypothesis of ALP-induced anomalies in gamma-ray transparency at GeV -- TeV energies.
\cite{Davies:2022wvj} investigated three bright, flaring flat-spectrum radio quasars using a refined jet model to explore the effects of ALPs on gamma-ray opacity. This work builds on earlier theoretical insights highlighting the significance of photon-photon dispersion in blazar jets~\cite{Davies:2021wqw}. Additionally, an analysis focusing on the cumulative effect in the extragalactic gamma-ray background was conducted by~\cite{Liang:2020roo}, which demonstrated consistency between gamma-ray data and expectations from EBL models.

At multi-TeV energies, however, challenges for EBL models have arisen. Detected on October 9, 2022, GRB 221009A, dubbed the ``Brightest of All Times'' (BOAT), set records for its gamma-ray luminosity and is estimated to occur once every 10,000 years \cite{Frederiks:2023bxg, Burns:2023oxn}. Observations with the James Webb Space Telescope linked it to a SN in a star-forming galaxy at redshift \( z = 0.151 \) \cite{Blanchard:2023zls}. Notably, it is the first burst with confirmed gamma-ray emission above 10 TeV. The LHAASO collaboration reported a 13 TeV gamma-ray event \cite{2022ATel15669....1D, LHAASO:2023lkv}, and the Carpet-2 array detected a 251 TeV event \cite{2022ATel15669....1D}. These high-energy detections raise questions about the universe's transparency and the role of ALPs.
Gao et al. \cite{Gao:2023und} analysed LHAASO data, exploring photon-ALP oscillations to explain gamma-ray transparency. Galanti et al. \cite{Galanti:2022chk} included Carpet-2 data, suggesting ALPs could explain the signal with adjusted mixing parameters. \cite{Carenza:2022kjt} found that ALPs could not fully account for both LHAASO and Carpet-2 events. \cite{Baktash:2022gnf} concluded that ALP explanations are plausible with modifications to mixing conditions.

\item \textit{Diffuse emission signals.}  
ALPs can generate large-scale diffuse emissions through various mechanisms. Any sufficiently bright, magnetized astrophysical source -- such as AGNs, SNe, or star-forming galaxies -- can act as an ALP source. The cumulative ALP emission from a given source class depends on its number density and redshift evolution. For instance, a diffuse axion background from cc-SNe was initially proposed in \cite{Raffelt:2011ft} and later extended to ALPs~\cite{Calore:2020tjw}. 
These ALPs can partially convert into gamma rays in the Galactic magnetic field via the Primakoff effect, imparting a spatial morphology to the ALP-induced gamma-ray emission that traces the Milky Way’s magnetic field structure. For a detailed spectral and spatial analysis of this diffuse background, see \cite{Calore:2021hhn}.
High-energy facilities have also been instrumental in probing ALP signatures in diffuse emission. \cite{Eckner:2022rwf} analysed the cumulative sub-PeV emission from star-forming galaxies due to photo-hadronic interactions and ALP conversion, identifying a potential exotic diffuse component, constrained using HAWC and Tibet AS$\gamma$ Galactic plane measurements. Similarly, \cite{Mastrototaro:2022kpt} extended this analysis to early diffuse data from LHAASO, offering further insights into the potential high-energy ALP contributions.
\end{itemize}

\begin{tcolorbox}[colback=pos!5!white, colframe=pos!75!black, title=Exercise 3]
 Derivation of constraints on ALP-photon coupling from the 
 gamma-ray observations of NGC 1275, cf.~\ref{sec:NGC1275_tutorial}.
\end{tcolorbox}

\section{WISP dark matter searches}
\label{sec:darkmatter}

\subsection{General considerations about dark matter}

DM is a cornerstone of modern cosmology, accounting for approximately 27\% of the total energy density of the universe. Despite extensive evidence from astrophysical and cosmological observations, its fundamental nature remains elusive. Various theoretical models propose DM candidates with a wide range of masses, interactions, and lifetimes, constrained by both direct and indirect experimental bounds.

For light DM candidates, with masses below 1 MeV, the only kinematically allowed decay channel is into two photons as already mentioned in the context of ALPs and sterile neutrinos in Sec.~\ref{sec:introPheno-decay}.
For non-relativistic particles, the energy density is determined by the DM mass and the particle number density. Since the DM particles at \( z = 0 \) are effectively at rest in galactic halos, any decay process would produce signals at the energy scale corresponding to the DM mass, offering a sharp spectral feature at half the mass of the DM particle.
On the other hand, contributions from all galaxies in the universe, redshifted and integrated over the star-formation history, can enhance the extragalactic gamma-ray background with a smoother contribution.  

A key element to predict the final photon flux expected from decay is the DM distribution in the target of interest.
Determining the DM distribution in galaxies requires a detailed analysis of mass modelling and local measurements. Orbital velocities of stars and gas, measured as a function of radial distance from the galactic centre, provide estimates of the gravitational potential. Comparing these measurements with expectations based on visible matter (disc, bulge, stars, and gas, including the galactic bar) often reveals the need for an additional unseen mass component, attributed to DM. 

The master equation for the expected DM decay flux per unit energy is:
\begin{equation}
    \left(\frac{d\Phi_\gamma}{dE} \right)_{\rm decay}= \frac{\Gamma(\rm DM \rightarrow \gamma \gamma)}{4 \pi m_a}\left(\frac{\mathrm{d}N_\gamma}{\mathrm{d}E}\right)_{\rm decay} \times \int_{\rm l.o.s.} \rho_{\rm DM}(\ell) \mathrm{d}\ell \, ,
\end{equation}
where $\Gamma(\rm DM \rightarrow \gamma \gamma)$ denotes the decay rate of DM into photons, $\left(\mathrm{d}N_\gamma/\mathrm{d}E\right)_{\rm decay}$ refers to the differential photon yield per energy per decay event and $\rho_{\rm DM}$ is the DM density profile of the target under scrutiny. The integral is performed along the line of sight (l.o.s.) between the observer and the target.

In the following section, we report some recent results in the search of ALPs decaying DM.
We remind the reader that, to constitute the cosmological DM abundance, the ALPs should be stable over cosmological timescales, meaning that their lifetime should be longer than the age of the universe. 
The condition $\tau_{a \rightarrow \gamma \gamma} \leq 13.8$ Gyr,     
imposes a minimal relation between the ALP-photon coupling and the mass, which ALPs DM should satisfy.

\subsection{Best targets for ALPs dark matter decay searches}
The indirect search for DM decay relies on identifying targets that are both rich in DM and exhibit minimal astrophysical backgrounds. The goal is to maximize the signal-to-noise ratio through optimised observation strategies and precise modelling of astrophysical contributions.

\textbf{Satellite galaxies of the Milky Way.} The Milky Way is surrounded by numerous satellite galaxies, many of which have substantial total masses (including both baryonic and DM components), making them promising targets for searches for decaying ALPs. The expected signal from ALP decay scales with the mass enclosed within the search volume. Studies such as \cite{Caputo:2018vmy, Caputo:2018ljp} have developed the theoretical framework for stimulated axion decay in various astrophysical environments, including Milky Way satellites, the Galactic centre, and galaxy clusters, with a focus on radio frequency signals.
Interest in satellite galaxies as targets for ALP searches has grown significantly in recent years. Early efforts, such as those by \cite{Blout:2000uc}, employed the 37-meter Haystack Observatory radio telescope to search for nearly monochromatic microwave photons from Local Group dwarf galaxies. Optical spectroscopy with the MUSE instrument was utilized in pioneering work by \cite{Regis:2020fhw} and further refined in \cite{Todarello:2023hdk}, which analysed data from five dwarf spheroidal galaxies of the Milky Way, including both classical and ultra-faint types, to constrain ALPs with masses around the eV scale. 
Other approaches have explored new observational tools for ALP studies. For example, \cite{Yin:2023uwf} investigated the potential of the SUBARU telescope to enhance constraints based on similar targets. More recently, \cite{Guo:2024oqo} used a two-hour radio observation of Coma Berenices conducted with the Five-hundred-meter Aperture Spherical radio Telescope (FAST) to place limits on the ALP parameter space. Although these constraints are weaker than those obtained with the CAST helioscope, they highlight the ongoing efforts to probe ALP properties through diverse observational techniques.

\textbf{Dark matter halo of the Milky Way.} While the Milky Way's DM halo is not technically an extragalactic object, its substantial mass makes it a compelling target for searches for ALP decay. Using blank-sky X-ray observations from XMM-Newton, \cite{Foster:2021ngm} investigated ALP decays in the Galactic halo. This effort was updated in \cite{Dessert:2023vyl} with data from Hitomi and projections for future observations with XRISM. Additional studies have leveraged eROSITA data \cite{Dekker:2021bos} and INTEGRAL data, the latter extending into the hard X-ray regime and focusing on the Galactic center \cite{Calore:2022pks}. 
In the infrared band, \cite{Roy:2023omw} presented a forecast of the James Webb Space Telescope's sensitivity to ALP decays within the Galactic halo. Meanwhile, \cite{Janish:2023kvi} utilized blank-sky observations from James Webb, originally intended for sky subtraction, to constrain ALPs with masses in the 0.8 -- 3 eV range. These efforts demonstrate the versatility of multi-wavelength approaches in exploring the ALP parameter space.

\textbf{Galaxy clusters.} Galaxy clusters, being the most massive virialized structures in the universe, serve as prime targets for detecting signals from decaying DM. \cite{Battye:2019aco} analyzed radio observations of the Virgo cluster to investigate spontaneous ALP decay. Similarly, \cite{Chan:2021gjl} examined radio data from the Bullet Cluster (1E 0657-55.8), achieving constraints on the ALP parameter space that are more stringent than those set by the CAST helioscope.

\textbf{Large-scale structure of the universe.} The filaments of dark and baryonic matter, which form the large-scale structure of the universe, could also be key to detecting decaying ALPs. The EBL has been utilized in various studies to probe ALPs \cite{Bernal:2022xyi}. For instance, \cite{Caputo:2020msf} focused on the near-infrared background angular power spectrum, including emissions from galaxies and intra-halo light, to constrain an additional component arising from ALP decay by comparing it with data from the Hubble Space Telescope and \textit{Spitzer}. \cite{Libanore:2024hmq} showed that future broadband ultraviolet surveys, such as GALEX or the upcoming ULTRASAT satellite, could tighten current ALP limits by focusing on the ultraviolet segment of the EBL. The optical component of the EBL was examined in \cite{Nakayama:2022jza} and \cite{Carenza:2023qxh} to investigate potential links between ALP decay and anisotropy in the COB flux, especially regarding the excess uncovered by the \textit{New Horizon} space probe's Long Range Reconnaisance Imager (LORRI). This excess, observed in LORRI images, reveals a discrepancy with expected COB flux, exceeding galaxy-based estimates by roughly a factor of two \cite{Bernal:2022wsu}. Additionally, \cite{Shirasaki:2021yrp} constrained eV-mass ALPs by cross-correlating line intensity and weak lensing maps, using optical and infrared data. By analyzing the cosmic photon background across various wavelengths, \cite{Porras-Bedmar:2024uql} provided constraints on a broad ALP mass range, also challenging interpretations of the LORRI excess in terms of ALPs.
Distant galaxies, such as quasars, can also be directly utilized in the search for ALPs. \cite{Wang:2023imi} investigated the decay of ALP DM by stacking dark-sky spectra from the Dark Energy Spectroscopic Instrument (DESI) at the redshift of nearby galaxies, using data from DESI’s Bright Galaxy and Luminous Red Galaxy samples. \cite{Sun:2023wqq} demonstrated how discrepancies in current datasets of high-redshift quasars could be addressed by considering the presence of ALPs.

\section{Hands-on tutorial on axion-like particle physics}
\label{sec:tutorials}

The tutorial session featured an exercise block targeting more practical applications of the physics of ALPs covering the derivation of the fundamental consequences of axions and ALPs for standard electrodynamics and a gamma-ray data-analysis part. The relevant questions to be answered in this context were:
\begin{enumerate}
    \item How does the presence of ALPs coupling to the electromagnetic field strength tensor change the well-known Maxwell equations of electrodynamics?
    \item Based on the modified Maxwell equations (and a bit of higher-order Quantum Electrodynamics), what is the probability of photons converting into ALPs and vice versa in an external magnetic field?
    \item How do we experimentally detect ALP-photon conversions in the gamma-ray spectra of extragalactic objects like NGC 1275?
\end{enumerate}

The scenario we are studying here is that of so-called ``axion electrodynamics'', which has interesting consequences and can be derived from the principles of classical field theory. The Lagrangian of the electromagnetic field, $\mathcal{L} = -\frac{1}{4}F_{\mu\nu}F^{\mu\nu}$, yields the equations of motion of electrodynamics -- also known as the Maxwell equations --
\begin{equation}
\partial_{\mu}F^{\mu\nu} = 0
\end{equation}
in the absence of source terms for the electromagnetic field. Recall that the field strength tensor is defined as
\begin{equation}
F_{\mu\nu} = \partial_{\mu}A_{\nu} - \partial_{\nu}A_{\mu}
\end{equation}
with $A_{\mu} = (-\phi, \vec{A})$ being the four-vector of the electromagnetic potential. In fact, the equations of motion above only yield two of Maxwell's equations. The other two follow from a similar equation involving the dual field strength tensor $\tilde{F}_{\mu\nu}$ defined by
\begin{equation}
    \tilde{F}_{\mu\nu} = \varepsilon_{\mu\nu\rho\kappa}F^{\rho\kappa}
\end{equation}
and leading to:
\begin{equation}
\label{eq:2nd_maxwell}
\partial_{\mu}\tilde{F}^{\mu\nu} = 0.
\end{equation}
In the presence of an axion-like particle $a$, we can minimally couple it to the electromagnetic field strength tensor via
\begin{equation}
\label{eq:lagrangian_ALPED}
\mathcal{L} = -\frac{1}{4}F_{\mu\nu}F^{\mu\nu} - \frac{1}{4}g_{a\gamma}F_{\mu\nu}\tilde{F}^{\mu\nu}a - \frac{1}{2}\partial_{\mu}a\partial^{\mu}a - \frac{1}{2}m_a^2a^2,
\end{equation}
while the last two contributions are the kinetic and mass terms for the ALP field, respectively.

\subsection{Exercise 1: Derivation of Maxwell’s equations for axion electrodynamics}
\label{sec:tutorial_ex1}

\begin{tcolorbox}[colback=pos!5!white, colframe=pos!75!black, title=Exercise 1]
Derive the equations of motion of axion electrodynamics (i.e.~with respect to $A_{\mu}$) from the Lagrangian above and transform them into their formulation in terms of $\vec{E}$ and $\vec{B}$. To this end, recall the explicit form of the field strength tensor and its dual:

$$
F^{\mu\nu} = \left(\begin{array}{cccc}
0 & -E^1 & -E^2 & -E^3 \\
E^1 & 0 & -B^3 & B^2 \\
E^2 & B^3 & 0& -B^1 \\
E^3 & -B^2 & B^1 & 0 \\
\end{array}\right),\quad
\tilde{F}^{\mu\nu} = \left(\begin{array}{cccc}
0 & -B^1 & -B^2 & -B^3 \\
B^1 & 0 & E^3 & -E^2 \\
B^2 & -E^3 & 0& E^1 \\
B^3 & E^2 & -E^1 & 0 \\
\end{array}\right).
$$
\end{tcolorbox}

\paragraph{Solution.} To solve the exercise, we start with the Euler-Lagrange equations for classical field theory for the electromagnetic potential $A_{\mu}$:
\begin{equation}
    \partial_{\nu}\frac{\partial \mathcal{L}}{\partial (\partial_{\nu}A_{\mu})} - \frac{\partial \mathcal{L}}{\partial A_{\mu}} = 0
\end{equation}
As the Lagrangian does not explicitly depend on $A_{\mu}$, we need to compute:
\begin{equation}
\label{eq:eqm_deriv}
    \partial_{\nu}\frac{\partial \mathcal{L}}{\partial (\partial_{\nu}A_{\mu})}  =-\frac{1}{4}\partial_{\nu} \left[\frac{\partial F_{\kappa\lambda}}{\partial (\partial_{\nu}A_{\mu})}F^{\kappa\lambda}+F_{\kappa\lambda}\frac{\partial F^{\kappa\lambda}}{\partial (\partial_{\nu}A_{\mu})}+g_{a\gamma}\frac{\partial F_{\kappa\lambda}}{\partial (\partial_{\nu}A_{\mu})}\tilde{F}^{\kappa\lambda} a +g_{a\gamma}F_{\kappa\lambda}\frac{\partial \tilde{F}^{\kappa\lambda}}{\partial (\partial_{\nu}A_{\mu})} a\right]
\end{equation}
The appearing derivatives are all of the same kind -- besides some manipulations regarding lowering and raising indices -- so it suffices to calculate it for one example:
\begin{align*}
\frac{\partial F_{\kappa\lambda}}{\partial (\partial_{\nu}A_{\mu})} &= \frac{\partial_{\kappa}A_{\lambda}}{\partial (\partial_{\nu}A_{\mu})} - \frac{ \partial_{\lambda}A_{\kappa}}{\partial (\partial_{\nu}A_{\mu})}\\
 &= \delta^{\nu}_{\kappa}\delta^{\mu}_{\lambda} - \delta^{\nu}_{\lambda}\delta^{\mu}_{\kappa}\mathrm{.}
\end{align*}
If we plug this result (and its corresponding variations for the other derivatives) in Eq.~\ref{eq:eqm_deriv}, we arrive at:
\begin{align*}
    \partial_{\nu}\frac{\partial \mathcal{L}}{\partial (\partial_{\nu}A_{\mu})}  &=-\frac{1}{4}\partial_{\nu} \left[2(F^{\nu\mu} - F^{\mu\nu})+2g_{a\gamma}(\tilde{F}^{\nu\mu}-\tilde{F}^{\mu\nu}) a\right]\\
    &=-\frac{1}{4}\partial_{\nu} \left[4F^{\nu\mu} +4g_{a\gamma}\tilde{F}^{\nu\mu} a\right]\\
    &=-\partial_{\nu}F^{\nu\mu} - g_{a\gamma}(\partial_{\nu}a)\tilde{F}^{\nu\mu} \overset{!}{=} 0.
\end{align*}
The second line follows from the fact that both, $F^{\mu\nu}$ and $F^{\mu\nu}$, are antisymmetric tensors, thus, transposing them leads to a sign flip. In the third line, the contribution of the term involving $\partial_{\nu}\tilde{F}^{\nu\mu}$ drops since the second set of Maxwell equations (see Eq.~\ref{eq:2nd_maxwell}) still holds unchanged.

This yields a system of four equations, which need to hold in parallel. Let us separate the time component from the spatial components of the system as they lead to independent modified Maxwell equations. 
\paragraph{Case 1: $\mu = 0$.} Given the explicit expressions for the field strength tensor and its dual, we immediately read off that:
\begin{align}
\label{eq:axionED_1}
    0 &= -\partial_iE^i-g_{a\gamma}(\partial_ia)B^i\nonumber\\
    &=\nabla\cdot\vec{E} + g_{a\gamma} \nabla a \cdot\vec{B}\nonumber\\
    &\Leftrightarrow\nonumber\\
    \nabla\cdot\vec{E} &= -g_{a\gamma} \nabla a \cdot\vec{B}\mathrm{,}
\end{align}
where we used the usual vector analysis notation to denote divergence and gradient and the convention to refer to spatial indices with Roman letters.

\paragraph{Case 2: $\mu = 1,2,3$.} Again, from the explicit form of the involved field strength tensors we find that:
\begin{align*}
    \mu = 1&: 0 = \partial_tE^1-\partial_2B^3+\partial_3B^2-g_{a\gamma}\left[-(\partial_ta)B^1-(\partial_2a)E^3+(\partial_3a)E^2\right]\nonumber\\
    \mu = 2&: 0 = \partial_tE^2+\partial_1B^3-\partial_3B^1-g_{a\gamma}\left[-(\partial_ta)B^2+(\partial_1a)E^3-(\partial_3a)E^1\right]\nonumber\\
    \mu = 3&: 0 = \partial_tE^3-\partial_1B^2+\partial_2B^1-g_{a\gamma}\left[-(\partial_ta)B^3-(\partial_1a)E^2+(\partial_2a)E^1\right]\nonumber\\
\end{align*}
These three equations can be summarised in one applying the definitions of gradient and curl to form the second equation of motion of axion electrodynamics.
\begin{align}
\label{eq:axionED_2}
    \nabla\times\vec{B} &= \partial_t \vec{E} + g_{a\gamma} (\partial_t a)\vec{B} + g_{a\gamma}\nabla a \times \vec{E}\mathrm{.}
\end{align}

\subsection{Exercise 2: Derivation of the conversion probability in constant magnetic field}
\label{sec:tutorial_ex2}

\begin{tcolorbox}[colback=pos!5!white, colframe=pos!75!black, title=Exercise 2]
Derive the equations of motion for the axion field $a$. Assume that there exists a static and uniform background magnetic field $\vec{B}_0 = (B_0, 0, 0)^T$ pointing in $\hat{x}$-direction. In an axion-universe, an electromagnetic wave is propagating in $\hat{z}$-direction with two polarisation states, $e_{\parallel} = \hat{x}$ and $e_{\perp} = \hat{y}$, parallel and transversal to the background magnetic field, respectively. The propagating wave can be seen as a perturbation of the background magnetic field so we write:

$$
\begin{cases}
a(z, t) & = \delta a(z, t)\\
\vec{B}(z, t) & = \vec{B}_0 + \delta \vec{B}(z, t)\\
\vec{E}(z, t) & = \delta \vec{E}(z, t).
\end{cases}
$$

Since the equations of motion regarding the background field are trivially true, derive the equations of motion for the perturbations by linearising the equations of axion electrodynamics. State the results in terms of the electromagnetic potentials rather than the electromagnetic fields. You may work in Weyl gauge, i.e. $\delta A^{0} = \delta\phi = 0$ and $\nabla\cdot\delta\vec{A} = 0$. Recall that electric and magnetic fields are derived from their potentials via

$$
\vec{E} = -\nabla\phi - \partial_t\vec{A},\\
\vec{B} = \nabla \times \vec{A}.
$$

Assume that the perturbations travel as plane waves

$$
\delta  A_{\parallel}(z, t) = ie^{i\omega(z-t)}A_{\parallel}(z),\\
\delta  a(z, t) = e^{i\omega(z-t)}a(z),
$$

where the phase factor for $A_{\parallel}(z)$ is just a convention. Derive a linearised coupled propagation equation for the system

$$
\psi(z):=\left(\begin{array}{c}
a(z)\\
 A_{\parallel}(z)
\end{array}\right)
$$

based on the linearised equations of motion for the perturbations. Solve the equation for an initial state comprised of only photons and derive the conversion probability of photons into axions.
\end{tcolorbox}

\paragraph{Solution.} To derive the conversion probability of ALPs into photons when subject to a magnetic field, we need to start from the equations of motion of the axion field. Hence, instead of solving the Euler-Lagrange equations for the electromagnetic potential $A_{\mu}$, we now compute them with respect to $a$:
\begin{align*}
\frac{\partial\mathcal{L}}{\partial a} &= -\frac{1}{4}g_{a\gamma}F_{\mu\nu}\tilde{F}^{\mu\nu}-m_a^2a\\
\frac{\partial\mathcal{L}}{\partial (\partial_{\mu}a)} &= -\partial_{\mu}a\mathrm{,}
\end{align*}
which leads to an inhomogeneous Klein-Gordon equation for the axion field using the fact that $F_{\mu\nu}\tilde{F}^{\mu\nu} = -4\vec{E}\cdot\vec{B}$:
\begin{equation}
\label{eq:axion_eom}
\square a + m_a^2 a = -g_{a\gamma}\vec{E}\cdot\vec{B}\mathrm{.}
\end{equation}

Now we move to the scenario of the exercise; an electromagnetic wave is propagating in $\hat{z}$-direction immersed in a homogeneous and uniform magnetic field $\vec{B}_0$ that points in the $\hat{x}$-direction. This propagating wave is a perturbation on top of the uniform background magnetic field. Hence, the dynamics of the system are governed by the sum of the wave and background field. Firstly, the dynamics of the axion field perturbation evolve according to
\begin{align*}
(\square + m_a^2) \delta a &= -g_{a\gamma}\delta\vec{E}\cdot(\vec{B}_0+\delta\vec{B})\\
&= -g_{a\gamma}\vec{B}_0\cdot\delta\vec{E} \mathrm{,}
\end{align*}
which follows from dropping all terms quadratic (and higher) in the perturbations. Plugging in the respective expressions for the electric and magnetic field strengths in terms of electric potential $\phi$ and vector potential $\vec{A}$, we can write in Weyl gauge
\begin{align}
(\square + m_a^2) \delta a &= g_{a\gamma} \vec{B}_0\cdot\partial_t\delta\vec{A}\nonumber\\
&= g_{a\gamma} B_0\partial_t\delta A_{\parallel}
\end{align}
since the background magnetic field is pointing in $\hat{x}$-direction leaving only the component of the vector potential $ A_{\parallel}$ parallel to it as contributing factor. Secondly, we address the modified Maxwell equations in this setting employing the Weyl gauge and dropping all terms quadratic in the perturbations. On one side, we obtain with Eq.~\ref{eq:axionED_1}
\begin{align}
\nabla\cdot\delta\vec{E} &= g_{a\gamma}(\nabla \delta a) (\vec{B}_0+\delta \vec{B})\nonumber\\
&\Longleftrightarrow\nonumber\\
\nabla \cdot (-\partial_t \delta\vec{A}) &=g_{a\gamma}(\nabla \delta a) \vec{B}_0\nonumber\\
&\Longleftrightarrow\nonumber\\
-\partial_t \nabla\cdot \delta\vec{A} &=g_{a\gamma}(\nabla \delta a) \vec{B}_0\nonumber\\
&\Longleftrightarrow\nonumber\\
0 &=g_{a\gamma}(\nabla \delta a) \vec{B}_0
\end{align}
which is trivially true since $\delta a$ only depends on $z$ and hence $(\nabla \delta a) \vec{B}_0 = 0$. On the other side, from Eq.~\ref{eq:axionED_2} we derive
\begin{align}
\label{eq:wave_eq_ALP}
-\partial_t \delta\vec{E} + \nabla\times(\vec{B}_0+\delta\vec{B}) &= g_{a\gamma} (\partial_t \delta a)(\vec{B}_0 + \delta\vec{B}) + g_{a\gamma}\nabla \delta a \times \delta\vec{E} \nonumber\\
&\Longleftrightarrow\nonumber\\
-\partial_t \delta\vec{E} + \nabla\times\delta\vec{B} &= g_{a\gamma} (\partial_t \delta a)\vec{B}_0\nonumber\\
&\Longleftrightarrow\nonumber\\
\partial^2_t \delta\vec{A} + \nabla\times\nabla\times\delta\vec{A} &= g_{a\gamma} (\partial_t \delta a)\vec{B}_0\nonumber\\
&\Longleftrightarrow\nonumber\\
\partial^2_t \delta\vec{A} - \triangle\delta\vec{A} = \square\delta\vec{A}&=g_{a\gamma}(\partial_t \delta a)\vec{B}_0
\end{align}
where in the last line we made use of the identity: $\nabla\times\nabla\times\vec{A} = \nabla(\nabla\cdot\vec{A}) - \triangle\vec{A}$. We can split Eq.~\ref{eq:wave_eq_ALP} into two components following the vector potential parallel and perpendicular to the background magnetic field. This yields:
\begin{align}
\label{eq:wave-eq-photon}
\square\delta A_{\parallel}&=g_{a\gamma}(\partial_t \delta a)B_0\\
\square\delta A_{\perp}&=0
\end{align}

What remains to be solved are the two wave equations for $\delta a$ and $A_{\parallel}$ which we will perform in the plane-wave ansatz. The third equation for $A_{\perp}$ decouples from the ALP field and is not of interest to us. Consequently, we start with the propagating ALP field $\delta  a(z, t) = e^{i\omega(z-t)}a(z)$. First, we will simplify the wave equation and, in particular, the spatial double derivative with respect to $z$:
\begin{align*}
\partial_z^2  \delta  a(z, t) = \partial_z^2 e^{i\omega(z-t)}a(z) &= \partial_z\left[i\omega e^{i\omega(z-t)}a(z)+e^{i\omega(z-t)}\partial_z a(z)\right]\\
&= e^{i\omega(z-t)}\left[-\omega^2+2i\omega\partial_z+\partial_z^2\right]a(z)\\
&\simeq e^{i\omega(z-t)}\left[-\omega^2+2i\omega\partial_z\right]a(z)\mathrm{,}
\end{align*}
where the last simplification is an approximation that the ALP amplitude is not strongly varying along its propagation direction, i.e.~$\omega^{-1}|\mathrm{d}\delta a/\mathrm{d}z|\ll 1$. We assume the same for $\delta A_{\perp}$. As a result, we can simplify Eqs.~\ref{eq:wave_eq_ALP} and~\ref{eq:wave-eq-photon} by also inserting the expressions in the plane-wave ansatz. The ALP wave equation becomes
\begin{align*}
    (\square + m_a^2) \delta a(z, t) &= (\partial_t^2-\partial_z^2+m_a^2)\delta a(z, t) = g_{a\gamma}B_0 \partial_t\delta A_{\parallel}(z, t)\\
    &\Longleftrightarrow\\
    e^{i\omega(z-t)}\left[-\omega^2-2i\omega\partial_z+\omega^2+m_a^2\right]a(z) &= g_{a\gamma}e^{i\omega(z-t)}(-i^2\omega) B_0A_{\parallel}(z)\\
    &\Longleftrightarrow\\
    i\partial_za(z) &= \frac{m_a^2}{2\omega}a(z)-\frac{1}{2}  g_{a\gamma}B_0 A_{\parallel}(z)
\end{align*}
Similarly, the parallel component of the photon field propagates according to
\begin{align*}
    \square\delta A_{\parallel} = (\partial_t^2 - \partial_z^2) \delta A_{\parallel}&=g_{a\gamma}(\partial_t \delta a)B_0\\
    \Longleftrightarrow\\
    e^{i\omega(z-t)}\left[-\omega^2-2i\omega\partial_z+\omega^2\right]iA_{\parallel}(z) &= -e^{i\omega(z-t)}g_{a\gamma}i\omega B_0a(z)\\
    \Longleftrightarrow\\
    i\partial_zA_{\parallel}(z) &= -\frac{1}{2} g_{a\gamma} B_0a(z)
\end{align*}
Both equations result in a coupled Schrödinger-type equation for the state $\vec{\Psi}(z) = (a(z), A_{\parallel}(z))^T$, which we may compactly write as:
\begin{equation}
\label{eq:schroedinger-eq}
    i\partial_z\vec{\Psi}(z) = \left(\begin{array}{cc}
        \Delta_a & \Delta_{a\gamma} \\
        \Delta_{a\gamma} & 0
    \end{array}\right)\vec{\Psi}(z)=:M\vec{\Psi}(z)\mathrm{,}
\end{equation}
where we defined the quantities $\Delta_a := \frac{m_a^2}{2\omega}$ and $\Delta_{a\gamma} := -\frac{1}{2}g_{a\gamma}B_0$. 

To solve the equation, we proceed in the usual way of making an exponential ansatz and decoupling both components by diagonalising the appearing matrix so that the general solution reads for a system of $N$ coupled components:
\begin{equation}
    \Psi_i(z) = \sum_{j=0}^N\mathcal{O}_{ij}\left[\mathcal{O}^{-1}\vec{\Psi}_i(0)\right]_je^{-i\lambda_iz}\mathrm{,}
\end{equation}
where $\{\lambda_i\}_{i=1}^N$ is the set of $N$ eigenvalues of the matrix $M$ while $\mathcal{O}$ denotes the matrix diagonalising $M$; i.e.~$\mathcal{O}^{-1}M\mathcal{O} = \mathrm{diag}(\lambda_1,\ldots, \lambda_N)$. In our specific case of a $2\times2$ matrix, it is clear that $\mathcal{O}\in SO(2)$ for which we can leverage the usual parametrisation
\begin{equation}
\mathcal{O} = \left(\begin{array}{cc}
    \cos{\theta} & -\sin{\theta} \\
    \sin{\theta} & \cos{\theta}
\end{array}\right)\rm{.}
\end{equation}
The eigenvalues of our matrix $M$ are
\begin{align*}
\det({M-\lambda\mathrm{Id})}\overset{!}{=}0&\Longleftrightarrow -\lambda(\Delta_a-\lambda)-\Delta_{a\gamma}^2=0\\
&\Rightarrow \lambda_{1/2} = \frac{1}{2}\left(\Delta_a\pm\sqrt{\Delta^2_a+(2\Delta_{a\gamma})^2}\right).
\end{align*}
Therefore, the most general solution to our propagation problem reads
\begin{align*}
a(z) &= \left[\cos^2\!{\theta}e^{-i\lambda_1z}+\sin^2\!{\theta}e^{-i\lambda_2z}\right]a(0) + \cos\!{\theta}\sin\!{\theta}\left[e^{-i\lambda_1z}-e^{-i\lambda_2z}\right]A_{\parallel}(0)\\
A_{\parallel}(z) &= \cos\!{\theta}\sin\!{\theta}\left[e^{-i\lambda_1z}-e^{-i\lambda_2z}\right]a(0) + \left[\cos^2\!{\theta}e^{-i\lambda_1z}+\sin^2\!{\theta}e^{-i\lambda_2z}\right]A_{\parallel}(0)\mathrm{.}
\end{align*}
The specific solution we are looking for is with respect to an initial photon-only state, i.e.~$a(0) = 0$ and $A_{\parallel}(0) = 1$. In this setup, we can evaluate the conversion probability $P_{\gamma\rightarrow a}(z)$ of photons into ALPs along the propagation direction by taking the modulus squared of the ALP amplitude $a(z)$. Consequently, we find that 
\begin{align}
    P_{\gamma\rightarrow a}(z) = |a(z)|^2 &= \cos^2\!{\theta}\sin^2\!{\theta}\left|e^{-i\lambda_1z}-e^{-i\lambda_2z}\right|^2\nonumber\\
    \Longleftrightarrow\nonumber\\
    &= 2\cos^2\!{\theta}\sin^2\!{\theta}(1-\cos\!\left[(\lambda_1-\lambda_2)z\right])\nonumber\\
    \Longleftrightarrow\nonumber\\
    &= 4 \cos^2\!{\theta}\sin^2\!{\theta}\sin^2\!\left(\frac{z\sqrt{\Delta^2_a+(2\Delta_{a\gamma})^2}}{2}\right)\nonumber\\
    \Longleftrightarrow\nonumber\\
    &= \sin^2\!{2\theta}\sin^2\!\left(\frac{z\sqrt{\Delta^2_a+(2\Delta_{a\gamma})^2}}{2}\right)
\end{align}
revealing the strong oscillation of the initial state $\vec{\Psi}(0)$ between photons an ALPs along the direction of propagation.

\subsection{Exercise 3: Derivation of constraints on ALP-photon coupling from the gamma-ray observations of NGC 1275}
\label{sec:NGC1275_tutorial}

As detailed in Sec.~\ref{sec:astro-HEP-ALP-signals}, ALPs may produce several different signatures in the gamma-ray emission of extragalactic objects. The third exercise of the tutorial was supposed to provide a hands-on introduction to those tools that facilitate the study of such phenomena in real gamma-ray data. Among the different gamma-ray experiments, the \textit{Fermi} LAT stands out, not only because of its large field of view and its revolutionising scientific discoveries but also because its entire collected data are publicly available.

As stated earlier, the LAT covers a wide energy range from a few tens of MeV to more than a TeV with a good energy resolution rendering the search for spectral irregularities in the spectra of blazars a feasible and promising objective. In this tutorial, we had a closer look at a study published by the \textit{Fermi}-LAT collaboration in 2016 \cite{Fermi-LAT:2016nkz}.

\paragraph{A very brief summary of the scope of \cite{Fermi-LAT:2016nkz}.} The authors selected the AGN NGC 1275 in the Perseus galaxy cluster to study its spectrum and the possibility of ALP imprints. They extracted the spectral distribution of NGC 1275 from 6 years of \textit{Fermi}-LAT data. They modelled the magnetic field inside the Perseus cluster as a homogeneous field with Gaussian turbulence to include the effect of axion-photon conversion inside the cluster while neglecting any conversion in the intergalactic medium. They performed a statistical analysis based on random realisations of the intracluster magnetic field of Perseus and Monte Carlo simulations of the \textit{Fermi}-LAT observations to search for a significant signal of ALPs. In lack of a clear signal, they set upper limits on the allowed space in the $(m_a, g_{a\gamma})$-plane finding stringent constraints on ALP-photon coupling for ALP masses between 0.5 and 5 neV.

\paragraph{Scope of the tutorial exercise.} The hands-on tutorial session during the COSMIC WISPers Training School was limited in time. Hence, we applied certain shortcuts to arrive at a final result. To reproduce the approach of the original paper, we would need much more time and computer power. However, the exercises reflect the actual rundown of a complete data analysis. 

In these lecture notes, we will not repeat each sub-task and solution with code examples. Instead, the full material for this tutorial part is provided on \texttt{zenodo.org} \cite{eckner_2025_14699436}. We refer the interested reader to this repository. Concretely, we covered the following topics and objectives:
\begin{itemize}
    \item Repeat the \textit{Fermi}-LAT data analysis of NGC 1275 according to the approach in the paper to extract its finely binned spectrum. The respective notebook is called \texttt{ngc1275\_fermipy.ipynb}. It contains an introduction to the \textit{Fermi} Science Tools and \texttt{fermipy} as well as the explicit application to NGC 1275. Note that during the tutorial session, this notebook was not shown due to time constraints.
    \item Brief introduction to \texttt{gammaALPs} \cite{me_manu_2021_4973513} to compute the photon survival probability given the magnetic field environment of the Perseus cluster used in \cite{Fermi-LAT:2016nkz}. This is the main part of the tutorial that was conducted during the session. The corresponding notebook is called \texttt{COSMICWISPers\_tutorial\_WISPs\_gammaray\_astrophysics\_solved.ipynb}.
    \item Setup the computation of the two theoretical expectations for the spectrum of NGC 1275: (i) without ALPs, (ii) with ALPs.
    \item Learn how to perform hypothesis testing with a $\chi^2$-likelihood function.
    \item Set some exclusion limits on ALP parameters.
\end{itemize}

\section*{Acknowledgements}

This article is based on the work from COST Action COSMIC WISPers CA21106, supported by COST (European Cooperation in Science and Technology).  
This publication is supported by the European Union's Horizon Europe research and innovation programme under the Marie Skłodowska-Curie Postdoctoral Fellowship Programme, SMASH co-funded under the grant agreement No. 101081355. The operation (SMASH project) is co-funded by the Republic of Slovenia and the European Union from the European Regional Development Fund.

\bibliographystyle{bib_style}
\bibliography{cost_school}

\end{document}